\documentclass[12pt]{article}
\usepackage{amsmath,amssymb,color}
\usepackage{cite}

\newread\epsffilein    % file to \read
\newif\ifepsffileok    % continue looking for the bounding box?
\newif\ifepsfbbfound   % success?
\newif\ifepsfverbose   % report what you're making?
\newif\ifepsfdraft     % use draft mode?
\newdimen\epsfxsize    % horizontal size after scaling
\newdimen\epsfysize    % vertical size after scaling
\newdimen\epsftsize    % horizontal size before scaling
\newdimen\epsfrsize    % vertical size before scaling
\newdimen\epsftmp      % register for arithmetic manipulation
\newdimen\pspoints     % conversion factor
\def\epsfbox#1{\global\def\epsfllx{72}\global\def\epsflly{72}%
   \global\def\epsfurx{540}\global\def\epsfury{720}%
   \def\lbracket{[}\def\testit{#1}\ifx\testit\lbracket
   \let\next=\epsfgetlitbb\else\let\next=\epsfnormal\fi\next{#1}}%
\def\epsfgetlitbb#1#2 #3 #4 #5]#6{\epsfgrab #2 #3 #4 #5 .\\%
   \epsfsetgraph{#6}}%
\def\epsfnormal#1{\epsfgetbb{#1}\epsfsetgraph{#1}}%
\def\epsfgetbb#1{%

\openin\epsffilein=#1 \ifeof\epsffilein\errmessage{I couldn't open
#1, will ignore it}\else

   {\epsffileoktrue \chardef\other=12
    \def\do##1{\catcode`##1=\other}\dospecials \catcode`\ =10
    \loop
       \read\epsffilein to \epsffileline
       \ifeof\epsffilein\epsffileokfalse\else

          \expandafter\epsfaux\epsffileline:. \\%
       \fi
   \ifepsffileok\repeat
   \ifepsfbbfound\else
    \ifepsfverbose\message{No bounding box comment in #1; using defaults}\fi\fi
   }\closein\epsffilein\fi}%

\def\epsfclipoff{\def\epsfclipstring{\ifepsfdraft\space clip\fi}}%
\epsfclipoff
\def\epsfsetgraph#1{%
   \epsfrsize=\epsfury\pspoints
   \advance\epsfrsize by-\epsflly\pspoints
   \epsftsize=\epsfurx\pspoints
   \advance\epsftsize by-\epsfllx\pspoints

   \epsfxsize\epsfsize\epsftsize\epsfrsize
   \ifnum\epsfxsize=0 \ifnum\epsfysize=0
      \epsfxsize=\epsftsize \epsfysize=\epsfrsize
      \epsfrsize=0pt

     \else\epsftmp=\epsftsize \divide\epsftmp\epsfrsize
       \epsfxsize=\epsfysize \multiply\epsfxsize\epsftmp
       \multiply\epsftmp\epsfrsize \advance\epsftsize-\epsftmp
       \epsftmp=\epsfysize
       \loop \advance\epsftsize\epsftsize \divide\epsftmp 2
       \ifnum\epsftmp>0
          \ifnum\epsftsize<\epsfrsize\else
             \advance\epsftsize-\epsfrsize \advance\epsfxsize\epsftmp \fi
       \repeat
       \epsfrsize=0pt
     \fi
   \else \ifnum\epsfysize=0
     \epsftmp=\epsfrsize \divide\epsftmp\epsftsize
     \epsfysize=\epsfxsize \multiply\epsfysize\epsftmp
     \multiply\epsftmp\epsftsize \advance\epsfrsize-\epsftmp
     \epsftmp=\epsfxsize
     \loop \advance\epsfrsize\epsfrsize \divide\epsftmp 2
     \ifnum\epsftmp>0
        \ifnum\epsfrsize<\epsftsize\else
           \advance\epsfrsize-\epsftsize \advance\epsfysize\epsftmp \fi
     \repeat
     \epsfrsize=0pt
    \else
     \epsfrsize=\epsfysize
    \fi
   \fi

   \ifepsfverbose\message{#1: width=\the\epsfxsize, height=\the\epsfysize}\fi
   \epsftmp=10\epsfxsize \divide\epsftmp\pspoints
   \vbox to\epsfysize{\vfil\hbox to\epsfxsize{%
      \ifnum\epsfrsize=0\relax
        \includegraphics{\ifepsfdraft}%
      \else
        \epsfrsize=10\epsfysize \divide\epsfrsize\pspoints
        \includegraphics{\ifepsfdraft}%
      \fi
      \hfil}}%
\global\epsfxsize=0pt\global\epsfysize=0pt}%

{\catcode`\%=12 \global\let\epsfpercent=%\global\def\epsfbblit{%BoundingBox}}%

\long\def\epsfaux#1#2:#3\\{\ifx#1\epsfpercent
   \def\testit{#2}\ifx\testit\epsfbblit
      \epsfgrab #3 . . . \\%
      \epsffileokfalse
      \global\epsfbbfoundtrue
   \fi\else\ifx#1\par\else\epsffileokfalse\fi\fi}%

\def\epsfempty{}%
\def\epsfgrab #1 #2 #3 #4 #5\\{%
\global\def\epsfllx{#1}\ifx\epsfllx\epsfempty
      \epsfgrab #2 #3 #4 #5 .\\\else
   \global\def\epsflly{#2}%
   \global\def\epsfurx{#3}\global\def\epsfury{#4}\fi}%

\def\epsfsize#1#2{\epsfxsize}

\let\epsffile=\epsfbox

\makeatletter \@addtoreset{equation}{section} \makeatother

\newcommand{\be}{\begin{equation}}
\newcommand{\ee}{\end{equation}}
\newcommand{\bea}{\begin{eqnarray}}
\newcommand{\eea}{\end{eqnarray}}
\def\del{\partial}

\def\a{{\alpha}}

\def\g{{\gamma}}

\def\ym{{\rm YM}}

\thispagestyle{empty}

\begin{document}
\begin{flushright}
{\bf hep-th/0608118}
\end{flushright}

\vspace{30pt}

\begin{center}

{\Large\bf No-Drag String Configurations\\ [1mm]
for Steadily Moving Quark-Antiquark Pairs\\ [3mm]
in a Thermal Bath}

\vspace{40pt}

{\large Philip C. Argyres, Mohammad Edalati,\\ [2mm]
and J.F. V\'azquez-Poritz}

\vspace{20pt}

{\it Physics Dept., Univ.\ of Cincinnati, Cincinnati OH 45221-0011}

\vspace{10pt}

\centerline{\tt argyres, edalati, jporitz@physics.uc.edu}

\vspace{40pt}

\centerline{{\bf{Abstract}}}
\end{center}

\noindent We investigate the behavior of stationary string
configurations on a five-dimensional AdS black hole background which
correspond to quark-antiquark pairs steadily moving in an ${\cal
N}=4$ super Yang-Mills thermal bath.  There are many branches of
solutions, depending on the quark velocity and separation as well as
on whether Euclidean or Lorentzian configurations are examined.

\newpage

\section{Introduction and summary}

The AdS/CFT correspondence provides a powerful tool with which to
study the strong-coupling behavior of certain non-Abelian gauge
theories in terms of semi-classical supergravity descriptions
\cite{malda9711, gkp9802, wit9802,agmoo}. The most-studied example
is four-dimensional ${\cal N}=4$ $SU(N)$ supersymmetric Yang-Mills
theory (SYM) which, in the limit of large $N$ and large 't Hooft
coupling, is described by type IIB supergravity on $AdS_5\times
S^5$.  Since at finite temperature the superconformal invariance of
this theory is broken, and since fundamental matter can be added by
introducing D7-branes \cite{kk0205}, it is thought that this model
may shed light on certain aspects of strongly-coupled QCD plasmas.
For example, for any strongly-coupled large $N$ gauge theory with a
gravity dual, the dimensionless ratio of the shear viscosity over
entropy density has been found to be $1/4\pi$ \cite{kss0309, bl0311,
kss0405, b0408} in rough agreement with some hydrodynamic models of 
RHIC collisions \cite{s0312,s0405}.

More generally, the RHIC experiment has raised the issue of how to
calculate the transport properties of relativistic partons in a hot,
strongly-coupled gauge theory plasma.   For example, one would like
to calculate the friction coefficient and jet quenching parameter
which are measures of the rate at which partons lose energy to the
surrounding plasma \cite{baier,kovner,gyulassy,jacobs}. Using
conventional quantum field theoretic tools one can calculate these
parameters only when the partons are interacting perturbatively with
the surrounding plasma. The AdS/CFT correspondence may be a suitable
framework in which to study strongly-coupled QCD-like plasmas.
In fact, attempts to use the AdS/CFT correspondence to
calculate these quantities have been made in \cite{lrw0605,
hkkky0605,gub0605,herzog2,st0605} and were generalized in various ways in
\cite{buch0605,h0605,cg0605,fgm0605,vp0605,sz0606,cg0606,lm0606,as0606,
Peeters:2006iu,aem0606,Gao:2006se,fgmp0607,Liu:2006nn,cag0607,mtw0607,
Caceres:2006ta}.

Finite-temperature ${\cal N}=4$ $SU(N)$ SYM theory is equivalent to
the near-horizon limit of type IIB supergravity on the background of
a large number $N$ of non-extremal D3-branes stacked at a point. From
the perspective of five-dimensional gauged supergravity, this is the
background of an AdS black hole whose Hawking temperature equals the
temperature of the gauge theory.
According to the AdS/CFT dictionary, string configurations on this
background can correspond to quarks and antiquarks in an ${\cal
N}=4$ SYM thermal bath \cite{ry9803,malda9803,rty9803,bisy9803},
where the quark bare mass is determined by the radial location of
the string endpoints on a probe D7-brane.

A stationary single quark can be described by a string that
stretches from the probe D7-brane to the black hole horizon.  The
semi-infinite string solution with a tail which drags behind a
steadily moving endpoint and asymptotically approaches the
horizon has been proposed \cite{hkkky0605,gub0605,herzog2} as the
configuration dual to a steadily moving quark in the ${\cal N}=4$
plasma, and was used to calculate the drag force on the quark.

A stationary quark-antiquark pair or ``meson", on the other hand,
corresponds to a string with both endpoints ending on the D7-brane 
\cite{ry9803,malda9803,kk0205}.  A class of such solutions, namely 
smooth, static solutions ($v=0$), have been used to calculate 
the inter-quark potential in SYM plasmas. Smooth, stationary 
solutions ($v\neq0$) for steadily moving quark pairs exist \cite{Liu:2006nn,cag0607,Caceres:2006ta,sfetsos,gubser2} 
but are not unique and do not ``drag" behind the string endpoints 
as in the single quark configuration.  This lack of drag has been
interpreted to mean that color-singlet mesons are invisible to the
SYM plasma and so experience no drag (to leading order in large $N$)
although the string shape is dependent on the velocity of the meson
with respect to the plasma.  (These timelike Lorentzian string
solutions are reviewed in section 3, below.)

On the other hand, a prescription for computing the jet quenching
parameter $\hat q$ using the lightlike limit of spacelike Lorentzian
configurations has been proposed in \cite{lrw0605}.\footnote{See
footnote 14 of \cite{Liu:2006nn}.  We thank H. Liu, K. Rajagopal, and
U. Wiedemann for correspondence clarifying this point.} In this paper,
we restrict ourselves to timelike Lorentzian and Euclidean
configurations, and address spacelike Lorentzian configurations
in \cite{aev0612}.

\paragraph{Summary.}

Concretely, we consider a smooth stationary string in the background of
an AdS$_5$ black hole with metric 
\be\label{metric}
ds_5^2=h_{\mu\nu} dx^{\mu} dx^{\nu}
=\eta\,{r^4-r_0^4\over r^2 R^2}\,dx_0^2+\frac{r^2}{R^2}
(dx_1^2+dx_2^2+dx_3^2)+ {r^2 R^2\over r^4-r_0^4} \,dr^2.
\ee
$R$ is the curvature radius of the AdS space, and the black hole
horizon is located at $r=r_0$.
Since we are interested in (timelike) Lorentzian as well as Euclidean
configurations, we have introduced the factor $\eta$; for Lorentzian
signature $\eta=-1$ while for Euclidean signature it equals $+1$.

We put the endpoints of the string on a probe D7-brane at radius
$r_7$.\footnote{The background (\ref{metric}) can be lifted to ten
dimensions on a 5-sphere, where it is the near-horizon geometry of a
stack of $N$ non-extremal D3-branes.  The probe D7-brane wraps an
$S^3$ inside the $S^5$ and fills the entire AdS$_5$ background down
to a minimal radius $r_7$
\cite{kk0205,beegk0306,mmt0605,ok0605,afjk(1)0605,afjk(2)0605}. We 
assume no
motion on the $S^5$, and so use the five-dimensional perspective.}
The classical dynamics of the string in this background is described
by the Nambu-Goto action \be\label{ngaction} S = {\eta\over2\pi\a'}
\int\! d\sigma d\tau\, \sqrt{\eta\,G}, \ee where
$G=\det(G_{\a\beta})$, $G_{\a\beta} =h_{\mu\nu} \del_\a X^\mu
\del_\beta X^\nu$ is the induced metric on the string worldsheet,
and $\del_\a:=\del/\del\xi^\a$, with $\xi^\a= \{\tau, \sigma\}$
being the worldsheet coordinates. Here $X^\mu$ run over the five
coordinates $\{x_0,x_1,x_2,x_3,r\}$.

The steady state of a quark-antiquark pair
with constant separation and moving with constant velocity
either perpendicular or parallel to the separation of
the quarks can be described (up to worldsheet reparameterizations)
by the worldsheet embedding
\bea\label{shape}
{}[v_\perp]: &\quad&
x_0 = \tau, \quad
x_1= v \tau, \qquad\ \,\,\,
x_2= \sigma, \quad
x_3= 0,\quad
r = r(\sigma), \nonumber\\ [1mm]
{}[v_{||}]: &\quad&
x_0 = \tau, \quad
x_1= v \tau + \sigma, \quad
x_2= 0, \quad
x_3= 0,\quad
r = r(\sigma),
\eea
where the first is for the velocity perpendicular to the quark
separation, and the second parallel to it.
In both cases we take boundary conditions
\be\label{bcs}
0\le \tau\le T,\qquad
-\frac{L}{2} \le\sigma\le\frac{L}{2}, \qquad
r(\pm L/2)= r_7,
\ee
with $r(\sigma)$ smooth.\footnote{The endpoints of a string on a 
D-brane satisfy Neumann boundary conditions in the directions along the 
D-brane, whereas the above boundary conditions are Dirichlet, constraining 
the string endpoints to lie along fixed worldlines a distance $L$ apart 
on the D7-brane.  The correct way to impose these boundary conditions is 
to turn on a worldvolume background $U(1)$ field strength on the D7-brane 
\cite{hkkky0605} to keep the string endpoints a distance $L$ apart.}

In section 2 we derive the equations of motion describing
configurations corresponding to quark-antiquark pairs, in both
Lorentzian as well as Euclidean signature, which are stationary and
smooth. As argued in \cite{cag0607}, smooth string configurations
cannot drag behind their endpoints as they dip down from the D7-brane. Among these no-drag configurations, we concentrate on two
simple geometries, where the common velocity of the quark pair is
either perpendicular or parallel to their separation.

We examine the timelike Lorentzian solutions of these equations in
some detail in section 3.  In the case that the meson velocity is
perpendicular to the quark separation $L$, we review the known
solutions.  Up to a maximum $L\le L_c(v)$ which decreases with
increasing velocity, there are two branches of Lorentzian solutions:
one ``long" whose radial turning point is closer to the horizon than
the ``short" solution.  For $L>L_c(v)$, quark-antiquark pairs only
exist as free states described by two disconnected strings.  These
two branches, as well as the complete screening length, have been
discussed in
\cite{Liu:2006nn,cag0607,Caceres:2006ta,sfetsos,gubser2}. The long
string solution which makes it closer to the horizon has been argued
to be unstable \cite{cag0607,gubser2} and presumably decays into the
shorter string configuration which shares the same boundary
conditions.   This is supported by the fact, shown in section 4,
that the energy of the Euclideanized version of the long string
configuration is greater than that of the short string.

When the meson velocity is parallel to the quark separation,
the long string configurations are not smooth, but instead develop
a cusp at the midpoint for a range of velocities.  The tip of the cusp
is lightlike.  On the other hand, the short string solution is always smooth.

In section 4 we examine the Euclidean string configurations in
detail.  Their main interest lies in the fact that their relative
thermodynamic stability can be determined by comparing their actions
(energies).  However, we find that not all Euclidean solutions have
Lorentzian counterparts.

Euclidean configurations for which the velocity is parallel
to their separation share some of the same
characteristics as the timelike Lorentzian configurations.
Namely, there tend to
be two different branches of solutions, except when
${v>1}$, for which there is only one.  (There is no restriction
to ${v<1}$ for Euclidean strings; $v$ is more properly thought
of as a slope parameter, rather than a velocity.) Also,
there is a maximum distance between the quark and the antiquark past
which only free quarks exist.

On the other hand, for Euclidean configurations with velocity
perpendicular to their separation, the discussion of the various
branches of string solutions becomes more involved. Firstly, some of
the branches no longer have a maximum value of $L$. Secondly, the
number of branches now depends on the velocity. In particular, for
low velocities there are actually four branches of solutions, while
only two branches exist for higher velocities. The solutions
in which the string dips closer to the black hole horizon are
less energetically favorable.

\section{Equations of motion}

According to the AdS/CFT correspondence \cite{agmoo} strings ending
on the D7-brane are equivalent to quarks in a thermal bath in
four-dimensional finite-temperature ${\cal N}{=}4$ $SU(N)$ super
Yang-Mills (SYM) theory.  The standard gauge/gravity dictionary is
that \be\label{dictionary} N=R^4/(4\pi\a'^2 g_s), \quad \lambda =
R^4/\a'^2, \quad \beta = \pi R^2/r_0, \quad m = r_7/(2\pi\a') ,
\ee where $g_s$ is the string coupling, $\lambda:=g_\ym^2 N$ is the
't Hooft coupling of the SYM theory, $\beta$ the inverse
temperature of the SYM thermal bath, and $m$ the quark mass at zero
temperature.  In the semiclassical string limit, {\it i.e.},
$g_s\to0$ and $N\to\infty$, the supergravity approximation in the
gauge/gravity correspondence holds when the curvatures are much
greater than the string length, $\ell_s := \sqrt{\a'}$.
%This means we trust the correspondence (without having to include
%$\a'$ corrections from string theory) when $(R,r_0,r_7)\gg\ell_s$,
%or, equivalently, when $(\lambda,\sqrt\lambda\ell_s/\beta,m\ell_s)\gg1$.
Furthermore, in this limit the mass of the quark is identified
with the energy (in some units) of the associated string
which, for a static configuration, is just proportional to the value of the Nambu-Goto action
(\ref{ngaction}).

Now, the string embeddings (\ref{shape}) considered here (and
elsewhere) depend on three additional parameters: $T$, $L$, and $v$.
According to the standard AdS/CFT dictionary, these are the time 
for which the quarks are propagating, their separation at a given 
time, and their common velocity, respectively-- all in the plasma 
rest frame.  Also, it is easy to see from (\ref{metric}) and (\ref{shape}) 
that, for Lorentzian ($\eta=-1$) solutions, the proper velocity $V$ 
of the string endpoints on the $r=r_7$ surface in the AdS black hole 
background is related to the velocity $v$ in the four-dimensional 
field theory by
\be\label{realV} 
V = {r_7^2 \over\sqrt{r_7^4 - r_0^4}}\, v . 
\ee
Note that real string solutions must have the same signature 
everywhere on the worldsheet.  Thus, a string worldsheet will be 
timelike or spacelike depending on whether $V$, rather than $v$, 
is greater or less than 1. In particular, the string worldsheet 
is timelike for $V<1$ which in terms of $v$ translates into 
$v<\sqrt{1-(r_0/r_7)^4}$.  (In Euclidean signature, $v$ is more
properly thought of as an angular parameter, and $V$ is likewise
a proper version of this parameter as measured in the Euclidean
AdS black hole metric.  This will be discussed more in section 4.2.)

With the embeddings (\ref{shape}) and boundary conditions
(\ref{bcs}), the action becomes
\bea
{}[v_\perp]: &\quad&
S=\frac{\eta T}{\g\,\pi\a'}
\int_0^{L/2} d\sigma \sqrt{{r^4-\g^2 r_0^4\over R^4}
+ {r^4-\g^2 r_0^4\over r^4-r_0^4} r'^2}, \nonumber\\
{}[v_{||}]: &\quad& S=\frac{\eta T}{\g\,\pi\a'} \int_0^{L/2} d\sigma
\sqrt{\g^2 {r^4-r_0^4\over R^4} + {r^4-\g^2 r_0^4\over r^4-r_0^4}
r'^2}, \eea where $r' := \del r/\del\sigma$ and 
\be
\g^2 := {1\over 1+\eta v^2} .
\ee
Thus, $\g$ is the usual relativistic $\g$-factor for
Lorentzian signature ($\eta=-1$), while it is always
a positive number less than 1 for Euclidean signature ($\eta=+1$).

We find for the equations of motion \bea\label{E} {}[v_\perp]:
&\quad& r'^2=\frac{1}{\g^2\,a^2 r_0^4 R^4}
(r^4-r_0^4) (r^4-\g^2 [1+a^2] r_0^4) , \nonumber\\
{}[v_{||}]: &\quad&
r'^2=\frac{\g^2}{a^2 r_0^4 R^4}
(r^4-r_0^4)^2 {(r^4-[1+a^2]r_0^4) \over (r^4-\g^2 r_0^4)} ,
\eea
where $a^2$ is a real integration constant.  Although we have
written $a^2$ as a square, it can be either positive or
negative.  Using (\ref{E}), the determinant of the induced
worldsheet metric becomes
\bea\label{G}
{}[v_\perp]: &\quad&
G=\eta\frac{1}{\g^4\,a^2 r_0^4 R^4}
(r^4-\g^2 r_0^4)^2, \nonumber\\
{}[v_{||}]: &\quad&
G=\eta\frac{1}{a^2 r_0^4 R^4} (r^4-r_0^4)^2 .
\eea
Thus, the sign of $G$ is the same as that of $\eta a^2$ (since
the other factors are squares of real quantities).
In particular, for Euclidean signature ($\eta=+1$) all real
worldsheets have $G>0$, and so we must take $a^2>0$.  For Lorentzian
signature ($\eta=-1$), the worldsheet is timelike ($G<0$) for $a^2>0$ and
spacelike for $a^2<0$.

The reality of $r'$ implies that the right sides of (\ref{E})
must be positive in all these different cases.  This positivity
then implies certain allowed ranges of $r$.  There can only be
real string solutions when the ends of the string, at $r=r_7$,
are within this range.  The edges of this
range are (typically) the possible turning points, $r_t$, for the
string.  We will describe the possible values of $r_t$ for
the timelike Lorentzian and Euclidean cases in the following sections.

Given these turning points, (\ref{E}) can be integrated for a string
solution which goes from $z_7$ to the turning point $z_t$ and back
to give
\bea\label{L}
{}[v_\perp]: &\quad&
L/\beta=\frac{2a\g}{\pi} \int_{z_t}^{z_7}
{dz\over\sqrt{(z^4-1)(z^4-(1+a^2)\g^2)}}, \nonumber\\
{}[v_{||}]: &\quad&
L/\beta=\frac{2a}{\pi\g} \int_{z_t}^{z_7}
{dz\sqrt{z^4-\g^2}\over(z^4-1)\sqrt{z^4-(1+a^2)}},
\eea
where we have used $r_0=\pi R^2/\beta$. Also, in (\ref{L}) we have
rescaled $z=r/r_0$ and likewise $z_t:=r_t/r_0$ and $z_7:=r_7/r_0$.
%(The absolute value takes care of cases where $z_7<z_t$.)
These integral expressions determine the integration constant
$a^2$ in terms of $L/\beta$ and $v$.

Also, we can evaluate the action for the solutions of (\ref{E}) to be:
\bea\label{S}
{}[v_\perp]: &\quad&
S= \frac{\eta T\sqrt\lambda}{\g\beta}
\int_{z_t}^{z_7} {(z^4-\g^2)\, dz \over
\sqrt{(z^4-1)(z^4-\g^2[1+a^2])}}
%- \int_1^{z_7} dz\,\sqrt{\frac{(1+\eta\,v^2)z^4-1}{z^4-1}}
, \nonumber\\
{}[v_{||}]: &\quad&
S= \frac{\eta T\sqrt\lambda}{\g\beta}
\int_{z_t}^{z_7} dz \sqrt{z^4-\g^2 \over
z^4-[1+a^2]} ,
\eea
where we have used $R^2/\a'=\sqrt\lambda$.
%The plus or minus signs are to be chosen depending on the
%relative sizes of $z_7$, $z_t$, and $\g^2$, and will
%be discussed in specific cases below.
For finite $z_7$, these integrals are convergent.  They diverge
when $z_7\to\infty$, and need to be regularized by subtracting
the self-energy of the quark and the antiquark \cite{ry9803, malda9803}.
%This regularization will be discussed in more detail in section 4.

\section{Timelike Lorentzian solutions}

Turning now to timelike Lorentzian ($\eta=-1$) string configurations,
we see from (\ref{G}) that the integration constant $a^2$ must be
positive.  An analysis of (\ref{E}), bearing in mind (\ref{realV}),
easily shows that real solutions can exist only for ${v< \sqrt{1-
z_7^{-4}}}$ and as long as the string is at radii satisfying
\bea\label{tlradii}
{}[v_\perp]: &\quad&
r^4/r_0^4\ >\ \g^2(1+a^2), \nonumber\\ [2.5mm]
{}[v_{||}]: &\quad&
r^4/r_0^4\ >\ \mbox{max}\left\{\g^2,\ 1+a^2\right\}.
\eea
We will first briefly review the case in which the
velocity of the quark-antiquark pair is perpendicular to their
separation \cite{Liu:2006nn,cag0607,
Caceres:2006ta} and then consider the parallel case.

\subsection{Timelike Lorentzian: perpendicular velocity}

Equation (\ref{tlradii}) implies that the radial turning point
of the string is at $z^4 := (r/r_0)^4 = \g^2(1+a^2)$.  It
also implies that, for a given velocity parameter $v$, the minimum 
D7-brane radius $z_7:= r_7/r_0$ reached by the probe must 
also be set to be greater than this value.  ($z_7$ should 
also be set greater than the critical value $z_7^c \approx 
1.02$, below which the D7-brane dips into the horizon, changing 
the topology of the space \cite{beegk0306, mmt0605, ok0605, 
afjk(1)0605, afjk(2)0605}.)

\begin{figure}[t]
   \epsfxsize=3.0in \centerline{\epsffile{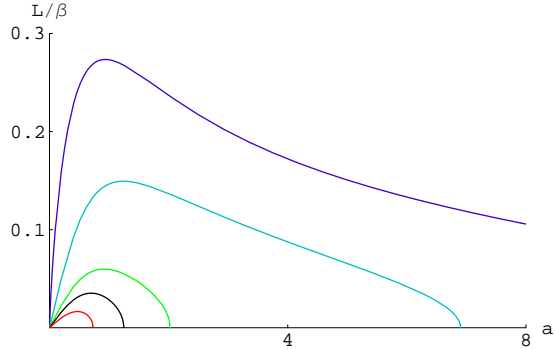}}
   \caption[FIG. \arabic{figure}.]{\footnotesize{$L/\beta$ as a function
   of $a$ for timelike Lorentzian string configurations with velocity
   perpendicular to the quark separation, $z_7=2$, and $\g=1$ (dark blue),
   $1.5$ (light blue), $2$ (green), $3$ (black), and $3.8$ (red).}}
   \label{fig1}
\end{figure}

For a given choice of $z_7$, (\ref{L}) can be numerically integrated
to give $L/\beta$ as a function of $a$, as shown in figure 1 for a
few sample velocities with the choice $z_7=2$.  (A similar plot has
already been presented in \cite{cag0607}.)
The qualitative features of the plots are not sensitive to the
particular value of $z_7$, though one should bear in mind that
the range of allowed velocity parameters $v$ depends on $z_7$,
since the string endpoints become spacelike for $\g>z_7^2$, as
can be seen from (\ref{realV}).

Figure 1 illustrates the fact that, for each value of $L$ that is
less than a critical value $L_c(v)$, there are two corresponding
values of $a$, and $L_c(v)$ decreases with increased velocity 
$v$.  We shall refer to the branch of string configurations
with smaller (larger) $a$ for a given $L$ as the long (short)
configurations.  For $L>L_c$ there is no connected string solution:
$L_c$ corresponds to a complete screening length past which quarks
and antiquarks only exist as free states \cite{Peeters:2006iu,
Liu:2006nn,cag0607,Caceres:2006ta}.

\begin{figure}[t]
   \epsfxsize=4.8in \centerline{\epsffile{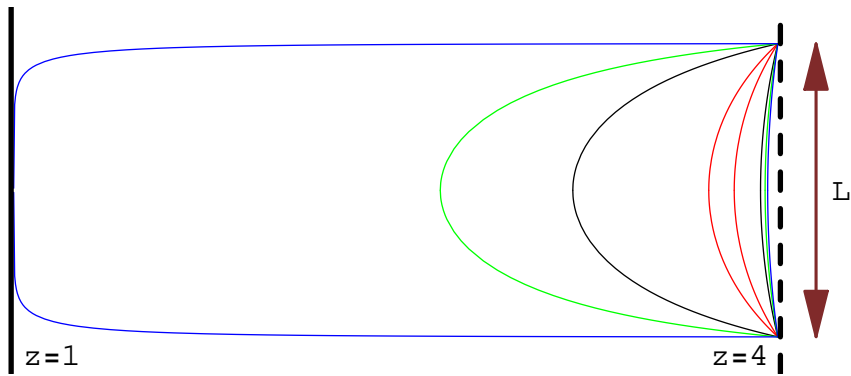}}
   \caption[FIG. \arabic{figure}.]{\footnotesize{Timelike Lorentzian
   string configurations with velocity perpendicular to the quark
   separation, $L/\beta= 0.1$, and $\g=1$ (blue), $1.5$ (green),
   $2$ (black), and $2.1$ (red).  For each velocity there are
   two string solutions, one long and one short.  The black hole horizon
   (solid black line) is at $z=1$ and the minimal radius reached by the
   probe D7-brane (dotted line) is at $z=z_7=2$.}}
   \label{fig2}
\end{figure}

We have plotted in figure 2 both long and short string configurations
for fixed $L/\beta$ and various velocities. The
radial direction is horizontal, the $x_2$-direction is vertical
and the velocity is orthogonal to both. The black hole horizon
is represented by the solid black line at $z=1$ and the probe
D7-brane corresponds to the dashed line at $z=z_7$.
For zero velocity (blue curves), the long string configuration
almost touches the black hole horizon. As the velocity
is increased so the string worldsheets become more nearly lightlike 
($V\to1$, or $\g\to4$), the long and short string configurations
shorten and lengthen, respectively, and approach a common
limiting shape (between the red curves).  They coincide when the
velocity reaches some $\g=\g_c$ ($\g_c\approx2.112$ for the specific
values of $L/\beta$ and $z_7$ in the figure).  This is the value
of the velocity parameter where
$L_c=L$; for greater velocities there are no string solutions
for this given $L$ and $z_7$.  A general qualitative property
of these solutions is that for any fixed $L$ and $z_7$
there is no lightlike limit of these timelike string configurations: 
the limiting $L=L_c$ is reached before $V=1$.

We will see in section 4
that the Euclidean counterparts of the long string solutions are
not energetically favored.  This indicates that this branch of
solutions is not stable: a long string state presumably
decays into the corresponding short string which has the same
boundary conditions. The instability of the long string
configurations was hypothesized in \cite{cag0607} and has been argued
for dynamically in \cite{gubser2}.

\subsection{Timelike Lorentzian: parallel velocity}

We will now look at the situation for which the velocity is in the
same direction as the quark-antiquark separation.
Recall that reality of $r'$ and the equation of motion (\ref{E}) implied
(\ref{tlradii}); that is, the allowed region is $z^4 > \mbox{max}
\{\g^2,\ 1+a^2\}$.  Unlike the perpendicular case, this
boundary is not always a smooth turning point of the string.
In particular, (\ref{E}) implies that $r'=0$ when $z=1+a^2$,
which is a smooth turning point (the string reaches a minimum);
but $r'=\infty$ when $z^4=\g^2$.  This latter behavior signals
the development of a cusp at the string midpoint.
As we discussed in section 2, $z^4=\g^2$ is also the
place where the string worldsheet changes from timelike to
spacelike signature.  Since, by (\ref{G}), real string
solutions cannot change their worldsheet signature, we
conclude that whenever $\g^2>1+a^2$ this cusp is
unavoidable.\footnote{
Without this physical argument, one might imagine that
the $r'=\infty$ vertical tangent is a signal not of a cusp,
but just that the string solution should be extended to
include a smooth but self-intersecting closed loop.  The
worldsheet embedding (\ref{shape}) we have used does not
allow for this extension, and so one might think that the
cusp could be avoided by using a different embedding.
For example, instead of using a parameterization in which
$x_1=v\tau+\sigma$ as in (\ref{shape}), which forces the
string to vary monotonically in the $x_1$ direction, one
might use a different parameterization with, say, $r=\sigma$
and $x_1=v\tau+x(\sigma)$ for some undetermined function
$x(\sigma)$.  This would, in principle, allow the string
to cross itself and form a smooth loop.  However, reworking
our calculations in this alternative parametrization gives
equations of motion completely equivalent to (\ref{E}).
Thus, this possibility is not realized, and the cusps cannot
be avoided, in agreement with the physical argument.}

\begin{figure}[t]
   \epsfxsize=3.5in \centerline{\epsffile{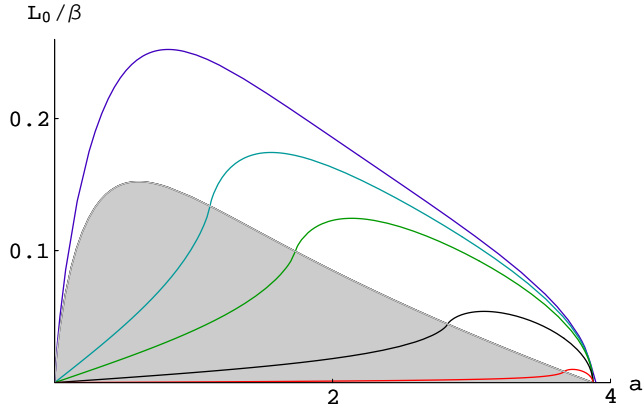}}
   \caption[FIG. \arabic{figure}.]{\footnotesize{$L_0/\beta$ as a
   function of $a$ for a timelike Lorentzian string with
   velocity parallel to quark separation, $z_7=2$, and $\g=1$
   (dark blue), $1.5$ (light blue), $2$ (green), $3$
   (black), and $3.8$ (red).  Strings corresponding to points
   in the shaded region have cusps.}}
   \label{fig3}
\end{figure}

It is not obvious when $\g^2>1+a^2$ is satisfied, since
$a^2$ is determined in terms of $v$ through (\ref{L}).
In figure 3 we integrate (\ref{L}) for various velocities
to give an indication of how $a$ depends on $L$ and $v$.
Figure 3 actually shows the quark separation $L_0:=\g L$
in the quark rest frame rather than the separation
$L$ in the plasma restframe.  There are two branches of 
solutions for $a$ when $L_0<L_{0c}(v)$, none
when $L_0>L_{0c}(v)$, and $L_{0c}(v)$ decreases for increasing $v$.
This is qualitatively similar to the perpendicular velocity case.

The shaded region in figure 3 is for $\g^2 > 1+a^2$, in which case the string solutions have a cusp.
Note that this region lies to the left of the maximum
of the constant-$v$ curves in the figure.  This means that the
large-$a$ (short string) solutions never have cusps but that,
depending on the values of $L_0$ and $v$, the long string solutions may.
Typically, for given $L_0$ the long string solutions for small
enough $v$ (close to $v=0$) and large enough $v$ (near where
$L_{0c}(v)$ approaches $L_0$) are smooth, while at intermediate
$v$ there are cusps.

\begin{figure}[t]
   \epsfxsize=5.5in \centerline{\epsffile{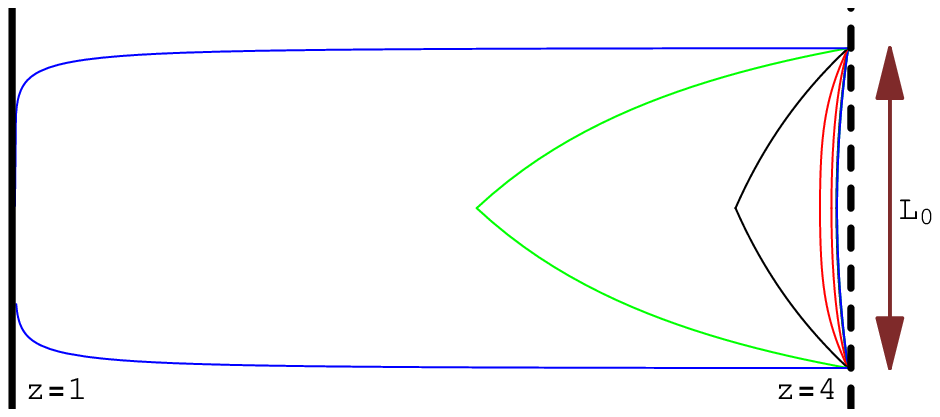}}
   \caption[FIG. \arabic{figure}.]{\footnotesize{Timelike
   Lorentzian configurations with velocity parallel to quark
   separation, $L_0/\beta= 0.1$, and $\g=1$ (blue),
   $1.5$ (green), $2$ (black) and $2.5$ (red).}}
   \label{fig4}
\end{figure}

This is illustrated for $z_7=2$ and $L_0/\beta=0.1$ in figure 4.
There the $\g=0$ (blue) and $\g=2.5$ (red) long strings have
no cusp, while the intermediate velocity (green and black) strings
do.\footnote{The appearance of a kink---for which there is a finite opening angle---rather
than a cusp in the green and black long strings in figure 4 is
misleading: the cusp behavior is apparent with sufficient resolution.}
Just as in
the case of perpendicular velocity, as $v$ is increased the long and
short strings approach one another until they coincide at
a critical value of the velocity parameter ($\g\approx2.6173$
for the values of the parameters in the figure), beyond which
there are no more connected solutions.  Note that this critical
velocity parameter is short of the lightlike worldsheet limit 
$\g=z_7^2$ so that, as in the case of perpendicular velocity, no lightlike 
worldsheet limit of the connected timelike configuration exists 
at fixed $L_0$ and $z_7$.

\section{Euclidean strings and their energetics}

Real Euclidean ($\eta=+1$) string configurations must have positive
integration constant $a^2$, by (\ref{G}).
An analysis of (\ref{E})
easily shows that real solutions can exist for any $v$
(since now $1\ge\g>0$ for all $v$) as
long as the string is at radii satisfying
\bea\label{eradii}
{}[v_\perp]: &\quad&
r^4/r_0^4\ >\ \mbox{max}\left\{1,\ \g^2(1+a^2)\right\},
\nonumber\\ [2.5mm]
{}[v_{||}]: &\quad&
r^4/r_0^4\ >\ 1+a^2.
\eea
Nothing special happens in Euclidean signature as the
``velocity" parameter $v\to1$.  Indeed, $v$ is more properly
thought of as an angular parameter in Euclidean space, though
we will still refer to it as the velocity parameter.

Note that there are Euclidean solutions which are not Wick rotations
of Lorentzian ones.
Lorentzian and Euclidean equations of motion (\ref{E})
are related to each other simply by taking $v^2\to -v^2$.
However, this does not mean that the corresponding solutions are simply related by 
Wick rotations since, under $v^2\to-v^2$, the
behavior of the turning points can change qualitatively.
In particular, timelike Lorentzian
solutions with perpendicular velocity always have $r_t^4=\g^2(1+a^2)r_0^4>r_0^4$ and so
the string never reaches
the horizon. On the other hand, for $a<v$ there is a branch of
Euclidean solutions which have the radial turning point on the black
hole horizon $r=r_0$.  This branch of solutions has no
physical Lorentzian counterpart.  Other examples of
Euclidean string configurations with no physical Lorentzian
counterpart are easy to come by.  For instance, the Wick rotation
of a steadily moving, purely radial Euclidean string stretched
between a probe D7-brane and the black hole horizon fails to exist
in Lorentzian signature, since there is an
intermediate radial point below which the string travels faster than
the speed of light.

\begin{figure}[h]
   \epsfxsize=4.8in \centerline{\epsffile{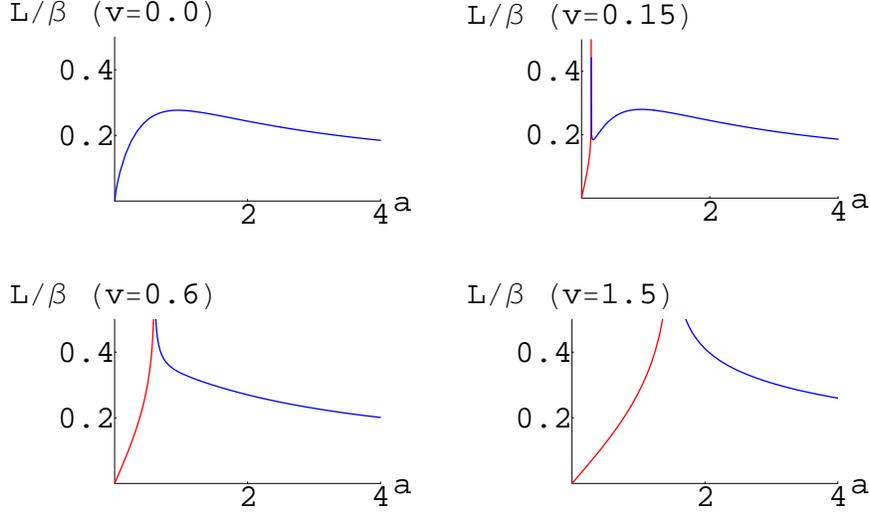}}
   \caption[FIG. \arabic{figure}.]{\footnotesize{$L/\beta$ as a
   function of $a$ with $z_7=2$ for Euclidean configurations with
   $v=0$, $0.25$, $0.5$ and $1$, for which the velocity is perpendicular
   to the quark separation.  The red and blue curves represent solutions
   with $a<v$ and $a>v$, respectively.}}
   \label{fig5}
\end{figure}

\subsection{Euclidean: perpendicular ``velocity"}

\begin{figure}[h]
   \epsfxsize=5.5in \centerline{\epsffile{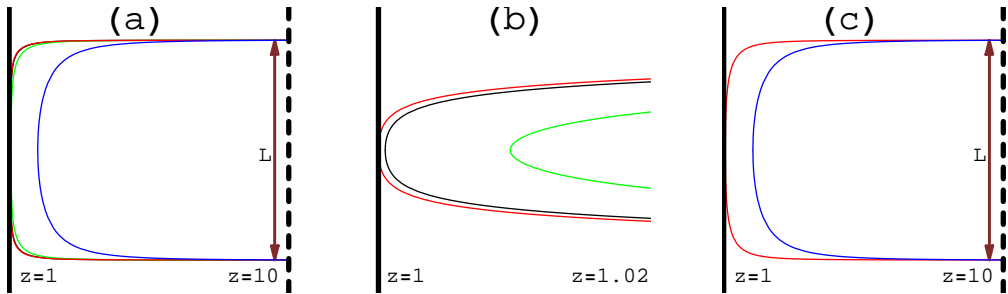}}
   \caption[FIG. \arabic{figure}.]{\footnotesize{Euclidean string
   configurations with perpendicular velocity, $L/\beta=0.25$ and
   $z_7=2$.  (a): Four solutions when $v=0.25$, with
   $a$ values of approximately $0.237$ (red), $0.290$ (black),
   $0.423$ (green) and $1.172$ (blue).  (b): Two solutions when
   $v=0.5$ with $a$ values approximately $0.397$ (red) and $1.503$
   (blue).}}
   \label{fig6}
\end{figure}

A numerical plot of $L/\beta$ as a function of $a$ for various velocities
is shown in figure 5. We have set $z_7=2$ as an example, though the
plot is qualitatively unchanged for other values of this parameter.
Since Wick rotation amounts to changing
the sign of $v^2$ in the equations, the configuration with $v=0$ is
exactly the same for Lorentzian and Euclidean signatures.  In
particular, there are no solutions for $L>L_c$, and for a given
value of $L<L_c$ there are two string configurations.  For $v>0$,
however, the story changes dramatically.  Firstly, there is no longer
a maximum value of $L$.  Secondly, the number of branches of solutions
depends on $L$ as well as the velocity.  For intermediate velocities,
two new branches of configurations emerge which have no Lorentzian
counterparts.  This is shown in the upper right region of figure 5
for $v=0.25$.  One new branch, which is denoted by the red curve, has
$a<v$ and exists for all values of $L/\beta$.  For small and large
values of $L/\beta$, there is only one branch of blue solutions but for
intermediate values of $L/\beta$ there are actually three branches.
For sufficiently large $v$, only one branch of blue solutions occurs.
This is illustrated in the lower left region of figure 5 for $v=0.5$.
For larger values of $v$ there are no qualitative changes, as
illustrated in the lower right of figure 5 for $v=1$.  (Nothing
special happens at $v=1$ in Euclidean signature.)

To better illustrate this, the four different string configurations
for $v=0.25$ and $L/\beta=0.25$ are plotted in figure 6(a).
Only the $a<v$ configuration, represented by the red curve, actually
touches the black hole horizon.  Only two of the branches of
configurations remain for all $L/\beta$ as the velocity is further
increased, as illustrated in figure 6(b) for $v=0.5$.

Which of these states is the
physical one for a given set of parameters can be determined by
comparing their energies.  The intuition that the blue
curve represents the energetically favorable solution, since it does
not stretch as far towards the black hole, is born out by a calculation
of the energies.
The energy of the Euclidean string configurations is given by
$S/T$, where $S$ is the Nambu-Goto action given by
(\ref{S}) and $T$ is the time interval.  

\begin{figure}[t]
   \epsfxsize=5.5in \centerline{\epsffile{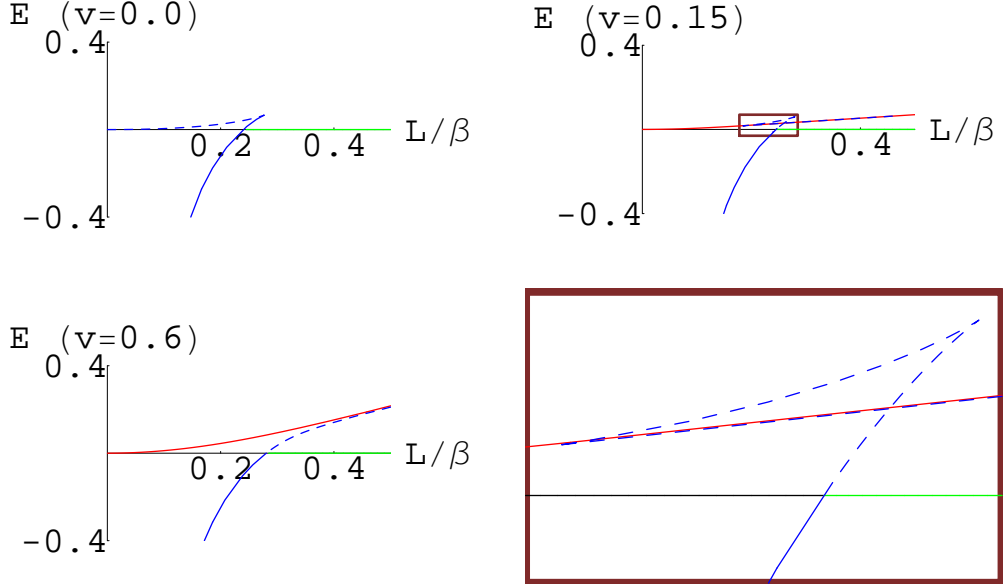}}
   \caption[FIG. \arabic{figure}.]{\footnotesize{Energy in units of
   $\sqrt{\lambda}/\beta$ versus $L/\beta$ for Euclidean
   configurations with perpendicular velocity, $z_7=2$, and $v=0$,
   $0.1$, and $0.5$.  The red and blue curves represent
   solutions with $a<v$ and $a>v$, respectively, while the green
   line is the subtracted energy of two straight strings.}}
   \label{fig7}
\end{figure}
\begin{figure}[t]
   \epsfxsize=3.5in \centerline{\epsffile{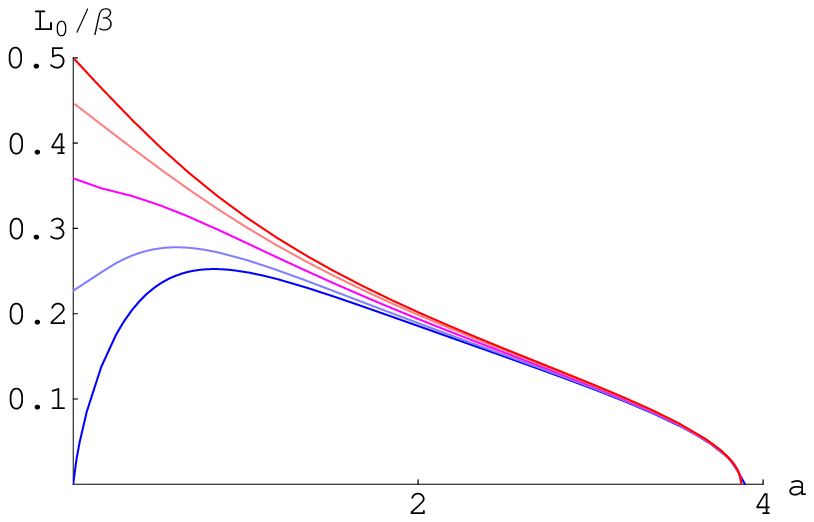}}
   \caption[FIG. \arabic{figure}.]{\footnotesize{$L_0/\beta$ as a function of
   $a$ for Euclidean string configurations with parallel velocity, $z_7=2$,
   and $v=0$ (blue), $0.5$ (light blue), $1$ (purple), $2$ (light red),
   and $5000$ (red).}}
   \label{fig8}
\end{figure}

It is more illuminating to
discuss the energy difference $E$ between these configurations
and some standard string configuration.  A simple and natural 
fiducial configuration to choose is that of two disconnected strings 
moving at ``velocity" $v$ which stretch from the probe D7-brane to 
the black hole horizon.   So in our discussion below we will measure 
energies in comparison to these straight string configurations, which
therefore have energy $E=0$ by definition.

$E\beta/\sqrt\lambda$ versus $L/\beta$ is plotted in figure 7 for
various velocities for the aforementioned configurations.   The
case of vanishing velocity has already been considered in
\cite{rty9803,bisy9803}.
As before, the red curve represents the string configuration with
$a<v$, which reaches the black hole horizon. There are
multiple configurations with $a>v$ for a given $L$ (blue curves),
depending on the velocity.  The energy of the fiducial straight string configuration
is given by the $E=0$ line.

As can be seen from figure 7, for $L$ less than a critical value,
the energetically favorable state is represented by the blue
curve.  This is the string configuration that remains the furthest
from the black hole horizon and is the Wick rotated counterpart of
the timelike Lorentzian short string solution with perpendicular
velocity that was discussed in section 3.  As the distance
between the quark and antiquark increases to the critical value,
the subtracted energy of this configuration becomes positive.
At this point, it is energetically favorable for the string to
separate into two straight strings (green line).  
Note that the long string configuration
(red curve) is always less energetically favorable than the
short strings (blue curve), which agrees with
the claim that the corresponding Lorentzian configurations are
unstable \cite{cag0607,gubser2}.

It is tempting to identify the transition from the short string
solution (blue) to the two straight string solution (green) as
the transition in the field theory from a bound quark and
antiquark pair to free quark pair due to complete screening
by the thermal bath.  However, this interpretation is problematical.
The reason is that, as mentioned earlier, the straight Euclidean
string is not the Euclidean rotation of any straight Lorentzian string 
solution (since at any nonzero $v$ such a straight Lorentzian string 
becomes lightlike before it reaches the horizon, and so fails to
exist as a solution).  A physically acceptable Lorentzian free quark 
solution is the dragged string solution of \cite{hkkky0605,gub0605},
so it may be more appropriate to compare the energy of the Euclidean short 
string solution (blue) to that of the Euclidean rotation of a pair
of dragged strings instead.  See \cite{cag0607} for a discussion of
the issues involved in making this comparison.  So the transition
between the blue and green configurations illustrated in figure 8
gives at best an upper bound on the critical $L$ at which complete
screening by the SYM thermal bath occurs.

\subsection{Euclidean: parallel ``velocity"}

For Euclidean string configurations with parallel ``velocity",
(\ref{eradii}) shows that the radial turning point is at
$r=(1+a^2)^{1/4}\,  r_0$.  Such solutions with ${V<1}$ are Wick
rotations of the timelike Lorentzian solutions with corresponding
turning point ({\it e.g.}, all those outside of the shaded region in
figure 3). In contrast to the configurations with perpendicular
velocity, there is always a maximum $L$ regardless of the magnitude
of the velocity. Also, there are no string configurations that reach
the black hole horizon. $L_0/\beta$ versus $a$ for various
velocities is shown in figure 8.  In Euclidean signature, $L_0:=\g L$
measures the shortest distance between the
``worldlines" of the endpoints of the strings. Since $\arctan(V)$
measures the angle between these worldlines and the constant-$x_1$
planes, in the limit $V\sim v\to\infty$ the worldlines coincide and
$L_0\to0$.  The curves in figure 8 ascend from $v=0$ to $v=\infty$.
For $V<1$ there are two solutions for each value of $L<L_c$.  The
short string configurations correspond to the part of the curves to
the right of the peak in figure 8, while the long configurations
correspond to the left side.  For $V>1$ there is only one solution and,
as $v\to\infty$, $L_c/\beta \to \sqrt{z_7^4-1}/(2z_7^2)$.  Thus, $L_c$
increases as the boundary worldlines are oriented more along the $x_1$
direction.

\vspace{.7cm}

%\centerline{\bf Acknowledgments}
\section*{Acknowledgments}

We would like to thank Jacques Distler, Paul Esposito, Joshua
Friess, Richard Gass, Steven Gubser, Andreas Karch, Juan Maldacena, 
Georgios Michalogiorgakis, Peter Moomaw, Leopoldo Pando Zayas, Silviu
Pufu and John Wittig for helpful conversations.
We are also indebted to Mariano Chernicoff, Antonio Garc\'ia,
Alberto G\"uijosa, Hong Liu, Krishna Rajagopal, Kostas Sfetsos,
and Urs Wiedemann for correspondence pointing out
errors and shortcomings of an earlier version of this paper.
We are grateful to the University of Michigan, the 2006 PITP school
at the IAS, and the 4th Simons workshop
at YITP for their hospitality.  This
research is supported by DOE grant FG02-84ER-40153.

\end{document}